\documentclass[journal,transmag]{IEEEtran}

\usepackage[pdftex]{graphicx}

\usepackage{url}
\usepackage[final]{changes}

\usepackage{tcolorbox}

\newcommand{\myparagraph}[1]{\vspace{1mm}\noindent{\bf #1}}

\begin{document}
%
\title{The Internet-of-Things Meets Business Process Management. \\A Manifesto}

%
\author{\IEEEauthorblockN{
 Christian Janiesch,
Agnes Koschmider,
Massimo Mecella,
Barbara Weber,\\
Andrea Burattin,
Claudio Di Ciccio,
Giancarlo Fortino,
Avigdor Gal,
Udo Kannengiesser,\\
Francesco Leotta,
Felix Mannhardt,
Andrea Marrella,
Jan Mendling,
Andreas Oberweis,\\
Manfred Reichert,
Stefanie Rinderle-Ma,
Estefania Serral Asensio,
WenZhan Song,
Jianwen Su,\\
Victoria Torres,
Matthias Weidlich,
Mathias Weske, and
Liang Zhang
}}

%



\IEEEtitleabstractindextext{%
\begin{abstract}
The Internet of Things (IoT) refers to a network of connected devices collecting and exchanging data over the Internet. These things can be artificial or natural and interact as autonomous agents forming a complex system. In turn, Business Process Management (BPM) was established to analyze, discover, design, implement, execute, monitor and evolve collaborative business processes within and across organizations. While the IoT and BPM have been regarded as separate topics in research and practice, we strongly believe that the management of IoT applications will strongly benefit from BPM concepts, methods and technologies on the one hand; on the other one, the IoT poses challenges that will require enhancements and extensions of the current state-of-the-art in the BPM field. In this paper, we question to what extent these two paradigms can be combined and we discuss the emerging challenges and intersections from a research and practitioner's point of view in terms of complex software systems development.
\end{abstract}

}

\maketitle

\IEEEdisplaynontitleabstractindextext

%
\IEEEpeerreviewmaketitle

\section{Introduction}
%
%
%
%

\IEEEPARstart{O}{ur} world is increasingly linked through a large number of connected devices, typically embedded in electrical/electronical components and equipped with sensors and actuators, that enable sensing, (re-)acting, collecting and exchanging data via various communication networks including the Internet: the Internet of Things \footnote{\added{Cf.  ITU: Internet of Things Global Standards Initiative, https://www.itu.int/en/ITU-T/gsi/iot/Pages/default.aspx}} \added{(see the dedicated box)}. As such, it enables continuous monitoring of phenomena based on sensing devices (wearable devices, beacons, smartphones, machine sensors, etc.) as well as analytics opportunities in smart environments (smart homes, connected cars, smart logistics, Industry 4.0, etc.) and the possibility to actuate feedback. Therefore, the IoT contributes to the recent trend known as big data, being one of the three main sources besides human sourced and process mediated data.

Business processes \added{(see the dedicated box)} represent a specific ordering of tasks and activities across time and place to serve a business goal, and often provides the driving force to system development. Process analytics, execution and monitoring based on IoT data can enable an even more comprehensive view of processes and realize unused potential for process optimization. As an example, in the past process analytics and in particular process mining has been hampered by the fact that processes are often incomplete or erroneous; with the IoT producing a large amount of data stored in the cloud \added{\cite{6742584}}, even more data become available for analysis, possibly resolving issues of incompleteness and enabling \added{the provision of}\deleted{providing} error correction methods based on multiple data items~\cite{batini2016data}.

In the literature, some works are emerging on combining Business Process Management (BPM) and IoT, e.g., utilizing sensor data to enable the actuation of services~\cite{janiesch2012beyond} or adapting running business processes to continuously align them with the state of the things (e.g., assets, humans, and machines)~\cite{marrella2018supporting}. Still, there are many open challenges to be tackled. Both BPM and IoT will benefit from a wider integration.

\begin{tcolorbox}[sharp corners, colback=cyan!30, colframe=blue!80!blue, title=The Internet of Things]
	
	The Internet of Things (IoT)~\added{\cite{ashton2009internet,10.1016/j.future.2013.01.010,holler2014machine}} is the inter-networking of physical objects (the things), being such things embedded systems with electronics hardware, software, sensors, actuators, and network connectivity. Such connected things collect and exchange data. Each thing is uniquely identifiable through its embedded computing system and is able to interoperate within the existing network infrastructure. While things act local, the IoT allows things to be controlled remotely across existing network infrastructures, including the Internet. 
	
	The interconnection of these smart objects/things \added{\cite{smartobjects2010}} is expected to usher in automation in nearly all fields. This creates opportunities for more direct integration of the physical world into computer-based and digitized systems, and results in improved efficiency, accuracy, and economic benefits besides increased automation and reduced human intervention. Experts estimate that the IoT will consist of about 30 billion objects \added{in}\deleted{by} 2020~\cite{nordrum2016popular}.
	
\end{tcolorbox}

\myparagraph{How IoT can benefit from BPM?} Let us consider a complex system with multiple components interacting within a smart environment being aware of the components' locations, movements, and interactions. Such a system can be a smart factory with autonomous robots, a retirement home with connected residents, or, at a larger scale, a smart city. While the parties in the system can track the movements of each component and also relate multiple components' behaviors to each other, they do not know the components' agendas. Often their interactions are based on \emph{habits}, i.e., routine \emph{low-level processes}, which represent \emph{recurring tasks}. Some of these routines are more time and cost critical than others, some may be dangerous or endanger others, and some may just be inefficient or superfluous. Knowing their agendas, their goals, and their procedures can enable a better basis for planning, execution, and safety.

\begin{tcolorbox}[sharp corners, colback=cyan!30, colframe=blue!80!blue, title=Business Process Management]
	
	Business Process Management is a well-established discipline that deals with the identification, discovery, analysis, (re-)design, implementation, execution, monitoring, and evolution of business processes~\cite{dumas2013fundamentals}. A business process is a collection of related events, activities, and decisions that involve a number of actors and resources and that collectively lead to an outcome that is of value for an organization or a customer. Examples of business processes include order-to-cash, procure-to-pay, application-to-approval, claim-to-settlement, or fault-to-resolution. To support business processes at an operational level, a BPM system (BPMS) can be used. As opposed to data- or function-centered information systems, a BPMS separates process logic from application code and, thus, provides an additional architectural layer. Typically, a BPMS provides generic services necessary for operational, software-enabled business process support, i.e., for process modeling, process execution, process monitoring, and user interaction (e.g., worklist management). When using a BPMS, software-enabled business processes are designed in a top-down manner, i.e., process logic is explicitly described in terms of a process model providing the schema for process execution. The BPMS is responsible for instantiating new process instances, for controlling their execution based on the process model, and for completing them. The progress of a process instance is typically monitored and traces of execution are stored in an event log and can be used for process mining~\cite{van2011process}, e.g., the discovery of a process model from the event log or for checking the compliance of the log with a given process model. 
	
	So far, the predominant paradigm to develop operational support for business processes has been based on the Model--Enact paradigm, where the business process has been depicted as a (graphical) process model, which then could be executed by a BPMS. This largely follows a top-down approach and is based on the idea of a central orchestrator that controls the execution of the business process, its data, and its resources. With the emergence of IoT, the existing Model--Enact paradigm is challenged by the Discover--Predict paradigm; it can be characterized as a bottom-up approach where data is generated from physical devices sensing their environment and producing raw events. Sensor data then must be aggregated and interpreted in order to detect activities that can be used as input for process mining algorithms supporting decision-making.
	
\end{tcolorbox}

\added{The solution of typical challenges in IoT, such as scalability -- massive number of devices, reliable coverage, power consumption problems -- energy harvesting and hardware/software optimization, can benefit as well by the knowledge of such agendas and goals. Finally, such a knowledge can support the design trade-offs involved in moving cloud services to edge of the network (the so called fog computing, i.e., defining the right allocation of where storing and processing data and offering services).}

\myparagraph{How BPM can benefit from IoT?} Let us consider a complex process with multiple parties interacting in the context of a business transaction. Such a process can be, for example, a procurement process, where goods are ordered, delivered, stored, and paid for. While the system can track each automatically-executed activity on its own, it relies on messages from other parties and manually entered data in the case of manual activities. If this data is not entered, or entered incorrectly, discrepancies between the digital (i.e., computerized representation of the) process and the real-world execution of the process occur. Similar concerns hold if the process participants do not obey the digital process under certain circumstances, e.g., an emergency in healthcare, or have not entered the data yet though in the real-world process the respective activity was already executed.
Such scenarios might be better manageable when closely linking the digital process with the physical world as enabled by the integration of IoT and BPM; e.g., the completion of manual activities can be made observable through usage of appropriate sensors \added{(e.g., \cite{DBLP:conf/caise/StertzMR17})}. IoT can complete BPM with continuous data sensing and physical actuation for improved decision making. Decisions in processes require relevant information as basis for making meaningful decisions. In general, it is not sufficient to retrieve this data solely from traditional repositories (e.g., databases and data warehouse\added{s}) providing historical data, but also up-to-date data are needed. Data from the IoT, such as events, provided through in-memory databases or complex event processing can be useful in this context. The IoT could reduce the need to manually signify the completion of manual tasks since sensor data is already available, leading to more accurate data, reduced errors, and efficiency gains.

In order to provide guidelines for system development, there still exist several challenges to be tackled. Particularly, it has to be understood:

\begin{itemize}
	\item how processes can improve the IoT by \emph{(i)} taking a process-oriented perspective and considering the process history to \emph{(ii)} bridge the abstraction gap between raw sensor data and higher-level knowledge extracted from this event data, and to \emph{(iii)} optimize the decision making in the large;
	\item how to exploit IoT for BPM by \emph{(i)} considering sensor data for automatically detecting the start and end of activities, \emph{(ii)} using event data for making decisions in a pre-defined process model, and \emph{(iii)} detecting discrepancies between the pre-defined model and actual enactment using event data for online process compliance checking and exception management.
\end{itemize}

In the following of this paper~\footnote{This paper has its roots in the Dagstuhl Seminar 16191 Fresh Approaches to Business Process Modeling, organized by Richard Hull, Agnes Koschmider, Hajo A. Reijers, and William Wong at the Leibniz Center for Informatics in Germany, May 8–13, 2016, cf. \url{http://drops.dagstuhl.de/opus/volltexte/2016/6696/}, to which many authors participated. Moreover, a preliminary version has been published on the Computing Research Repository (CoRR), abs/1709.03628, 2017, cf. \url{http://arxiv.org/abs/1709.03628}.}, taking these two general questions as starting point, we detail the key points in combining BPM and IoT and elaborate on benefits of BPM for IoT and IoT for BPM.

\begin{figure}
	\includegraphics[width=\columnwidth]{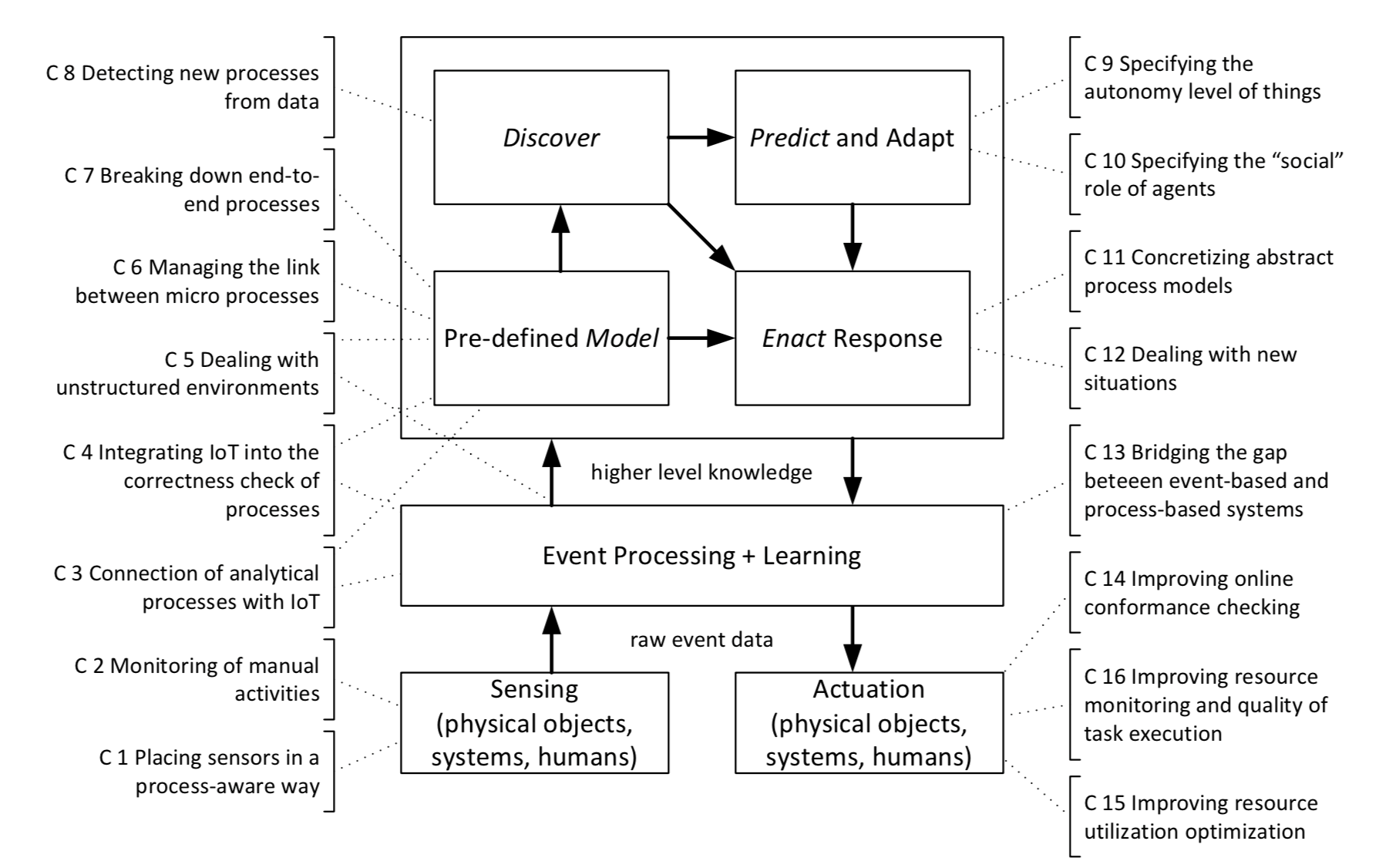}
	\caption{High-level overview showing the interaction between IoT and BPM. The numbering used in the blocks will correspond to the numbering of the different paragraphs in the text.}
	\label{fig:overview}
\end{figure}

\section{Intersections/challenges}

The IoT has to deal with a number of challenges; these include, for example, technological barriers such as computational limitations of embedded systems or the connectivity to back- end systems, security-related issues, \added{standardization and interoperability issues}, data privacy issues, untapped potential in data analytics, efficient methods for the organization of IoT systems, etc.~\cite{10.1016/j.future.2013.01.010}. The principal characteristic of the IoT is the communication between loosely-coupled objects, which mostly is accomplished asynchronously and ad-hoc. 

BPM deals with the discovery of models, the analysis of pre-defined models, the adaptation of models, and the enactment of business processes through software applications and systems. \deleted{We consider processes as explicit process representations (pre-defined models), which later are enacted.} Abstract processes can also be discovered from log files and suitable implementations for instantiation can be predicted. 

Accordingly, sensing and perception via sensors and decision based on sensors, as well as decision based on actuation according to individual goals/strategies, constitute fundamental tasks of the IoT. Thereby, sensing constitutes the input and actuation the output of any IoT-BPM interaction (see also Figure~\ref{fig:overview}). In between, raw event data are processed by event-based systems, transforming the input events to higher-level knowledge. In turn, the latter may be utilized by BPM concepts, methods or technologies to deal with the discovery of a \added{(process)} model, the analysis of a pre-defined model, the adaptation of a model and the enactment of \added{a model (of a } business process\added{)}. 

While the IoT generally focuses on communication and data flow, BPM approaches consider control flow, \deleted{big} process models \added{(large and ``in-the-large'')}, and synchronous interactions. In addition, most \added{of current} BPM approaches have \added{difficulties in}\deleted{troubles} dealing with non-routine, non-deterministic processes, whereas IoT applications typically involve these kinds of interactions.

Plenty of intersections, posing \added{new} challenges for researchers and practitioners, arise, as detailed in the following.

\subsection*{C1 – Placing the sensors in a process-aware way}

In order to collect all relevant data, sensors need to be carefully placed. It constitutes already a challenge to construct sensors and place them on agents (human or artificial) or in a smart \added{environment}\deleted{space}, such that they are non-intrusive but still efficient: sensors can be battery-less tags such as RFID, battery and renewable energy powered, or outlet-powered; and the communication methods can be wired or wireless. It is even more challenging to decide on the type of sensor and its placement with regard to its function in respect to the interaction between agents. A \added{(model of a}) business process may guide this placement since it offers knowledge about resources, locations and variants of behavior (enact\added{ment}), that need to be covered. As well, the trade-off between the cost of introducing additional sensing points and the expected increase in monitoring accuracy may be approached based on process knowledge.

\subsection*{C2 – Support for managing manually executed, physical processes}

In many \added{scenarios,}\deleted{settings, BPM approaches are used to automate} processes \added{are automated} through \deleted{the support of} a BPMS – Business Process Management System, in which some activities require the interplay between human operators and software/hardware modules; in many of these scenarios, there is an increasing use of mobile devices fostering the delivery of work items to the right users~\cite{pryss2016robust}. 

\deleted{In these settings,} Workers do not necessarily have to interact with the BPMS while carrying out physical tasks (e.g., moving boxes in a warehouse): sensors, which are connected to the BPMS, monitor whether or not such a task has started or ended. However, appropriate mappings from process activities to the \replaced{user interface}{GUI} and usable visualizations are needed \replaced{to allow}{allowing} actors (process participants) to perform their work in a natural way, without requiring non-value adding management tasks such as clicking on confirmation buttons.

\subsection*{C3 – Connection of analytical processes with IoT}

During process execution, a variety of information is required to make meaningful decisions. In turn, this information often needs to be available not only from traditional databases/data warehouses providing historical data, but it needs to be up-to- date and current. \deleted{In order to design systems providing such up-to-date information, and to judge the quality of the data analysis results from such applications,} It needs to be clear where the data stem from and where they have been used (\emph{data provenance}), as well as the overall quality \added{requirements} \deleted{of the data at-hand needs} to be ensured. \deleted{This is particularly critical for big data. Generally,} It becomes necessary to find a way to annotate the origin of data and use this (meta-)information in process models. So far, there is no \replaced{a widely accepted}{universal} \replaced{approach}{method} to connect the analytic processes of observation, analysis, and decision-making to business processes in a standardized way; recent attempts include the Decision Model and Notation (DMN) standard. Its focus, however, is on decision requirements, but less on the origin and use of decision data. Hence, it still needs to be investigated how to model quality and provenance in order to be exploitable at the process model level. 

Erroneous sensors, not working at all or delivering erroneous data, need to be discovered and excluded from any analysis. In turn, \deleted{this necessitates} a reasonable judgment on which sensor data might be erroneous \added{is needed:}\deleted{. Here,} the process context in which these data occur might be helpful to identify erroneous sensors as well as to cope with them.

\subsection*{C4 – Integrating the IoT with process correctness checks}

Well-known techniques for analyzing \deleted{the quality of} process models can contribute to improve the design of interactions in IoT, by finding deadlocks, livelocks, or dead activities \deleted{in respect to}\added{in} \deleted{the behavior or } interactions of \added{smart} objects. Deadlocks and livelocks are reasons why some processes may not terminate in the assumed time frame or not at all. While a rollback is a typical service in data management, it becomes much costlier and more complicated when managing processes and thus should be avoided. Dead activities do not harm a processes execution (unless they are supposed to be mandatory) since they will never be triggered. Yet, they represent a waste of resources as either or both, physical and/or virtual resources may have been reserved for these activities.

Therefore, designing correct process models which specifically consider the IoT nature of some components becomes crucial, as well as the verification of important properties.

\subsection*{C5 – Dealing with unstructured environments}

BPM offers a way to structure businesses. As such, it often assumes a controlled environment with a managed repository of versioned processes that can be orchestrated for the purpose of a single enterprise or be choreographed between parties in case of cross-organizational collaborations. Orchestration denominates the execution order of the interactions from the perspective and under control of a single party, whereas choreography describes public, i.e., globally visible, message exchanges, interaction rules and agreements made among multiple parties. Both concepts presume knowledge about the structure and/or interactions of each participating process. It is questionable whether orchestration and choreography still suffice as organizational concepts in an IoT world, which is much more ad hoc and situative (e.g., devices involved in the interaction might fail, deliver erroneous data, new devices may have to be flexibly added, etc.).

\subsection*{C6 – Managing the links between micro processes}

One approach to bridge the gap between IoT data and processes, would be to break end-to-end process models into micro processes representing habits and arrange them in a less prescriptive (control-flow) way. Modeling a small and possibly autonomous micro process does not necessarily require new modeling constructs or methods. Yet, the organization of hundreds/thousands of loosely coupled small processes may require new modeling constructs and methods to structure and represent their non-hierarchical interaction in human-readable form. 

Data-centric process paradigms offer promising perspectives in this context~\cite{steinau2019dalec}. For example, object-aware processes describe the behavior of single objects through micro processes, whereas the dynamic construction of linked objects as well as their synchronized execution is described and enforced through macro processes. However, respective approaches need to be enhanced to integrate physical objects as well as their behavior in the overall process.

\subsection*{C7 – Breaking down end-to-end processes}

\begin{figure}
	\includegraphics[width=\columnwidth]{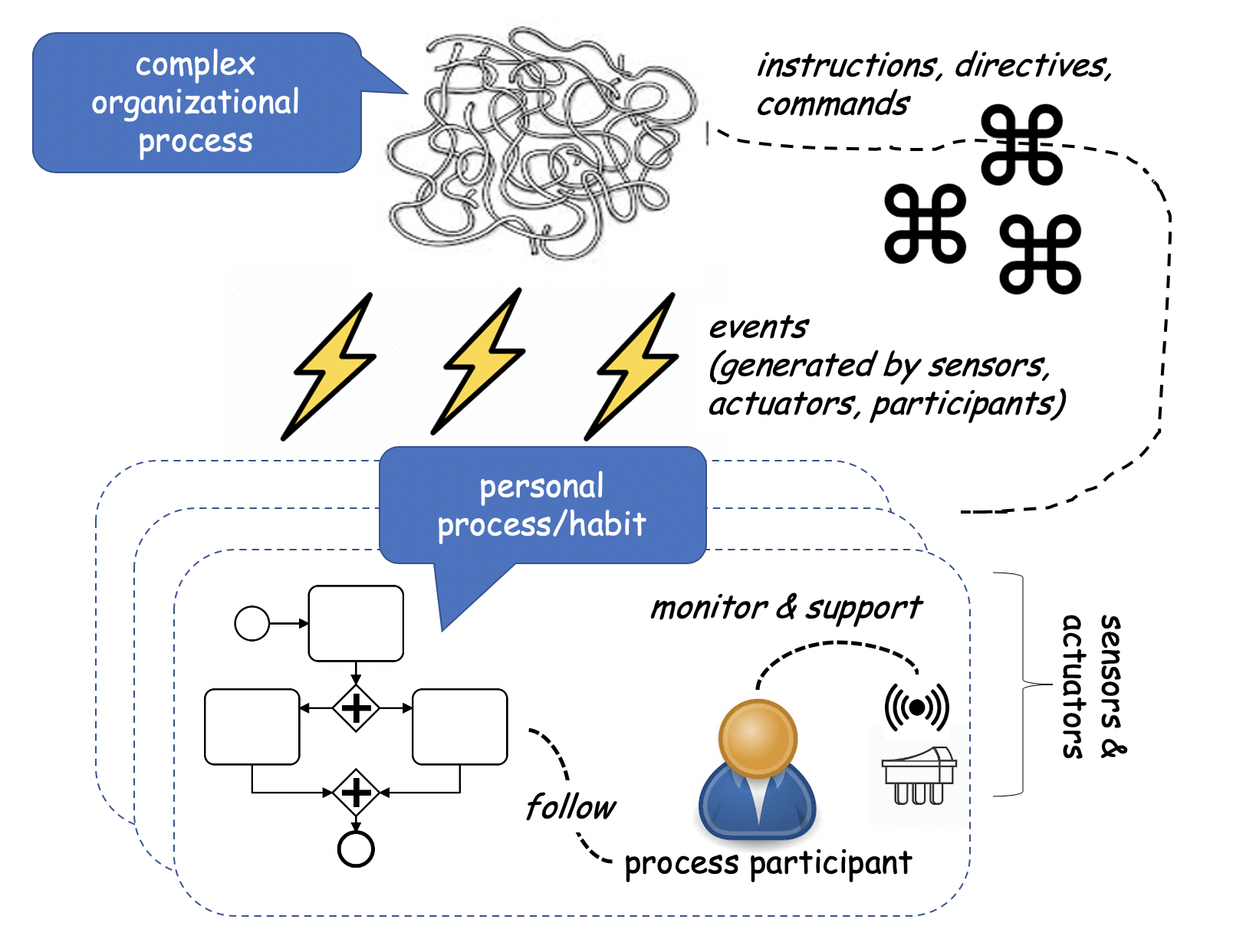}
	\caption{\added{The interplay of personal processes/habits wrt. complex organizational processes. Process participants follow habits, and are monitored and supported by sensors and actuators. Events and instructions/directives/commands are interconnecting the two layers, but without any rigid prescription and possibly through models to be dynamically mined.}}
	\label{fig:microprocesses}
\end{figure}

For a large class of processes (typically referred to as dynamic or knowledge-intensive~\cite{di2015knowledge}), the advent of overwhelming sensor data and things acting in the environment without central control but according to ``personal'' agendas, makes it practically impossible to define comprehensive end-to-end process models. Things will perform their own routines, so called repeated behaviour patterns or habits \added{(to be possibly mined, see \cite{Leotta2020})}. Accordingly, processes will have to be organized as \emph{event-driven micro processes} to represent these habits. Whereas the overall end-to-end business process itself may be modeled in traditional ways, the linking of micro-process models is far more complex; to cope with this emerging complexity, the possible interactions between micro-process models must not be described at the low level of message exchanges, but be put at a higher semantical level, similar to the utilization of semantic object relations for the purpose of object interactions in object-aware process management. \added{Figure \ref{fig:microprocesses} gives the intuition of such a complex interplay.}

\subsection*{C8 – Detecting new processes from data}

Designing a system in a bottom-up manner without prescriptive process models promises more flexible and inclusive processes. However, the question arises to what extent we can let the system just evolve and be discovered. When developing support for software-enabled business processes based on the principles of the IoT, an evolutionary self-organising process will take place in some respect. Thus, one must find the appropriate level of structuring and prescription without harming the capability to self-organize. There is a gap between IoT data and concepts at a model level to enable behavior prediction and to identify changes in behavior. The IoT allows deriving situational knowledge when tracking and evaluating data streams. Situational knowledge, in turn, is input to analyze prospective knowledge, which constitutes a dynamic task. Prospective knowledge addresses long-tail information about resources (e.g., how well is the person/thing doing? Are there any behavior changes expected?). Moreover, data streams from sensors need to be tracked, mapped to information entities, and simulated. Additionally, the output (goal) must be known (e.g., save time, save costs, improve health) and its derivation as well as the reconciliation of private goals must be mapped with organizational goals, which in turn is a challenge of the IoT. An alignment between event-based and process-oriented systems is indispensable in this context. A starting point could be to define goal-based deviation patterns and to provide modeling techniques considering sensor-data and event data.

\subsection*{C9 – Specifying the autonomy level of IoT things}

Objects in the IoT are able to react to events by executing tasks or entire processes. The execution of the latter \added{ones} is typically asynchronous and sometimes not explicitly started from a central coordinator. The execution of tasks or processes may further trigger certain reactions, for example the start of another process to correct deviating behavior. Yet, it is unfeasible to grant things full autonomy to decide everything without supervision. Hence, there has to be a concept of autonomy levels that dictate if things need supervision and may be vetoed, be it an individual or a group. Currently, there is no universal way to represent these levels of autonomy or to resolve conflicts originating from this distinction~\cite{parasuraman2000model,schillo2003taxonomy}. While different conceptualizations of individual and group autonomy exist, they have not been transferred to BPM \added{or IoT} yet. \deleted{Moreover, there is a lack of understanding on how to express them in a business process model, e.g. using patterns or further attributes.}

\subsection*{C10 – Specifying the roles of  things}

Organizations aim \replaced{at optimizing}{to optimize} their business processes based on organizational (i.e., group) goals. However, process participants often follow personal, i.e., individual processes or agendas with individual goals. The challenge is to synchronize/reconcile different, possibly conflicting goals. These agendas are typically mitigated through governance processes prescribing desired behavior. The individual goals of a thing are typically not precisely known or explicitly given. Furthermore, these processes may be less prescriptive micro processes or habits. Hence, holistic and prescritive governance may not be possible. Hence, it is an option to define and specify “social” behavior of things (such as self-interest, helpful, cooperative~\cite{kalenka1999socially}) to better coordinate and govern their behaviors. This becomes even more challenging \replaced{with}{when also considering robotics, i.e.,} the integration of human actors as well as robots in processes (raising issues like exchangeability, co-existence of different kinds of resources, etc.).

\subsection*{C11 – Concretizing abstract process models}

Abstract process models are sometimes used to model processes at design time without providing the details necessary for execution. This is a sensible approach when dealing with very dynamic scenarios. In these cases, it is possible to define the process, but the abstract model has to be turned into a concrete model later before being executable, for example by discovering available services as well as the conditions in which these services may be used. Context also includes physical data about users, e.g., location, devices the user carries with him (e.g., smartphone), etc. For the discovery phase \deleted{(see Figure~\ref{fig:overview})}, the semantics related to the services (i.e., what functionality can the service offer especially within the context of the process) should be available and it should be possible to reason over this for matchmaking purposes. In addition, the services' discovery phase may lead to changes in the schema of the original abstract process. Examples of corresponding changes include the skipping of certain tasks initially planned in the process or the addition of new fragments (e.g., combining two or more services either in sequence or parallel to achieve the task goal). \deleted{Despite all flexibility, this phase of instantiation presumes a given structure in form of the abstract process model.}

\subsection*{C12 – Dealing with new situations}

Individual ad hoc decisions may resolve a current situation from an individual's or a small group's point of view towards favorable results for them. In a complex business environment, foresightful and structured decision making cannot only achieve similar results but also save costs and time, and possibly improve the total quality. Deterministic event detection and correlation can be very well modeled and executed with event processing languages in complex event processing engines. However, the flexible discovery of new situations and the derivation of new responses constitute major technological challenges whose tackling can benefit from the combination with BPM. 

BPM methodologies and technologies can support the identification and selection of appropriate responses by recommending tasks, triggering tasks or whole processes, and automating as well as monitoring their execution. These reactions can be pre-defined using existing BPM technologies and learning can be based on the analysis of historic traces to identify beneficial habits from a higher-level perspective. Furthermore, reference models \added{\cite{Loos2007}} can help to identify state-of-the-art industry blueprints, which can be contextualized and instantiated to find a proper reaction for the context and the history of the situation. The capability of IoT sensing can be of additional benefit here. 

\subsection*{C13 – Bridging the gap between event-based and process-based systems}

A challenge is to bridge the gap between clouds of sensor data and event logs for process mining. Events captured by sensors are available in high volume, velocity, and variety. They are often affected by noise and errors. Process knowledge can be employed to support the identification of events from raw event data and in a subsequent step entire processes including their activities from event data. This is a non-trivial problem since event data belonging to different activities can be interleaving. Moreover, event data can belong to or be relevant for several activities, so that complex n:m relations between events and activities have to be considered. \deleted{The question of mapping start and end of event (streams) to the start and end of a process is closely related.} Once the activities have been discovered, the next challenge is to discover the corresponding processes, i.e., to correlate the activities with the corresponding process instances. Process knowledge and BPM methodologies \added{(e.g., \cite{DBLP:conf/caise/SenderovichRGMM16})} can support the discovery \deleted{of these habits}, the identification of the\deleted{ir} underlying interactions as processes as well as the optimization \deleted{of these habits} to reduce the waste of time and resources and \added{to} increase the safety of all involved agents. Process mining techniques provide promising ex post perspectives in this respect but require the presence of an event log that organizes the events in terms of traces representing the execution of a process instance. Similarly, but in an on-line fashion, complex event processing can be used to derive higher level knowledge from raw events to provide an ex-nunc perspective [2]. Here, the timely provisioning of events is crucial. 

\subsection*{C14 – Improving online conformance checking}

Conformance checking is a process mining technique that compares an existing process model with an event log of the same process. It can be used to check if the reality of process execution, as recorded in the log, conforms to the model and vice versa. Online conformance checking takes as input the context data and performs the comparison online. This requires high quality data and almost complete information. Again, the IoT as a data source and data management technolog\replaced{ies}{y} can play a major role and might improve the conformance checking of the actual physical execution with the execution order as recorded by the BPMS based on a secondary log of sensor data. Similarly, \added{IoT data can be used} for the checking and monitoring of compliance rules to be obeyed during process execution. 

\subsection*{C15 – Improving resource utilization optimization}

BPM can provide a governance structure for an organization, be it physical or virtual. BPM initiatives break up traditional functional silos and introduce process managers being responsible for processes across departments. While complex systems and the IoT is centered around situations to react to, BPM initiatives are organized around processes. This entails that some coordination instance responsible for priorities and resource provisioning can monitor and intervene with additional knowledge if necessary. In a pure IoT paradigm, there is the danger that decision will only produce local optima. The coordinating unit responsible for resource provisioning has advanced knowledge about the future behavior of agents since they have to follow their process models and, thus, can provide resources (e.g., computing power, network bandwidth, or things) with greater accuracy reducing processing time and thus increasing the throughput of a process. It also helps to reduce communication time-outs and thus, rollbacks, or abnormal process terminations \added{(cf. some initial results in \cite{10.1109/HICSS.2014.474, DBLP:journals/fgcs/0002JVWH15})}.

\subsection*{C16 – Improving resource monitoring and quality of task execution}

The execution of tasks in a business process consumes resources. These can be IT, such as storage capacity for process data, computing power for calculations in scientific workflows, artificial agents, such as as robots automatically executing manual tasks, or human beings entering or analyzing data or performing manual tasks. Also, machines, e.g., packing drugs, can be considered as resources (e.g., predictive monitoring, i.e., when does the machine have to be maintained taking its usage as well as historical data into account). 

All these resources might suffer from issues, which hinder optimal working conditions such as over- or under-utilization or even damage/ illness. IoT-based sensors can pick up these issues by measuring machine-behavior or human stress levels~\cite{adam2017design} and suggest changes to process execution to alleviate these effects. Furthermore, the IoT can support the execution of (knowledge-intensive) tasks in a process through context-specific knowledge provisioning, e.g., in terms of instructions or training materials on how to execute the task, or regulations that are relevant for the user's particular context. Sensor data can be leveraged to determine the actual context and to identify information needs (e.g., detection of cognitive overload or stress).

\section{Concluding remarks}

The IoT provides many opportunities for \replaced{organizations/companies/industries}{industry} as well as for personal use through the meaningful, yet dynamic interaction of humans, software, machines, and things. BPM is a well-established discipline that deals with the discovery, analysis, (re-)design, implementation, execution, monitoring, controlling and evolution of business processes. 

So far, both areas have been considered separately. In this paper we have formulated a number of points for the amalgamation of the IoT and BPM, which we deem important to be tackled in the near future in order for the IoT to benefit from business processes and vice-versa. 

\added{When adopting both IoT and BPM in the building of a complex system, we need to carefully consider the specific application scenarios, therefore the generalizability and adoption of practices, patterns, modeling approaches may be questionable. One of the challenging thesis of this paper is that general modeling, design and mining approaches should be devised in order to be able to consider different applications. An interesting preliminary question is how to classify, and according to which dimensions, IoT applications in order be able to perform such a generalization (cf. an initial study in \cite{DBLP:conf/bpm/MandalHOW18}). Also, not all scenarios for potential IoT application can equally benefit from BPM, e.g., the single app-controlled Phillips Hue lamp will not profit from BPM concepts, whereas a scenario that schedules maintenance appointments for a fleet of cars might.}

Before concluding, we would like to highlight a cross-issue, i.e., dealing with security and, in particular, privacy issues. \deleted{For example,} Privacy levels that exist at the sensors \replaced{layer}{level} might be different with respect to those at the BPM \replaced{one}{side}. A full-disclosure approach should be avoided, especially in contexts where sensitive (i.e., personal) information is collected. The most relevant challenge, in this case, is the communication between the two worlds, each of them with corresponding privacy/security levels and policies. The layer in charge of integrating these two sides should be designed according to the principles of privacy by design~\cite{langheinrich2001privacy}: ``identify and examine possible data protection problems when designing new technology and to incorporate privacy protection into the overall design, instead of having to come up with laborious and time-consuming ``patches'' later on''~\cite{schaar2010privacy}. This issue can also be seen as a ``non-functional requirement'' referring to C1, C3, C4, C6, C8, C13, and C14, but also others might be affected. Finally, partially related to the previous point, are ethical aspects of the integration of IoT and BPM: the introduction of raw events paves the way to a whole new set of analyses and explorations. On the one hand, these analyses must preserve the privacy of the individual (privacy is recognized as a fundamental right~\footnote{Cf. Article 8 of ``European Convention on Human Rights'', \url{http://www.echr.coe.int/Documents/Convention_ENG.pdf} and Article 12 of the ``Universal Declaration of Human Rights'', \url{http://www.ohchr.org/EN/UDHR/Documents/UDHR_ Translations/eng.pdf}.}). At the same time, the analyses should not be unfair and should not provide unequal treatment of people based on membership to a category or a minority. This problem is typically referred to as ``discrimination-aware data mining''~\cite{pedreshi2008discrimination}. More generally, the literature also talks about ``privacy-preserving data mining''~\cite{vaidya2006privacy}. There are several points that are directly affected by that such as C2-C6 and C13-C15. This is due to the set of analyses that the integration of IoT and BPM will make possible.


%



\ifCLASSOPTIONcaptionsoff
  \newpage
\fi



%


\newpage
%
\begin{IEEEbiography}[{\includegraphics[width=1in,height=1.25in,clip,keepaspectratio]{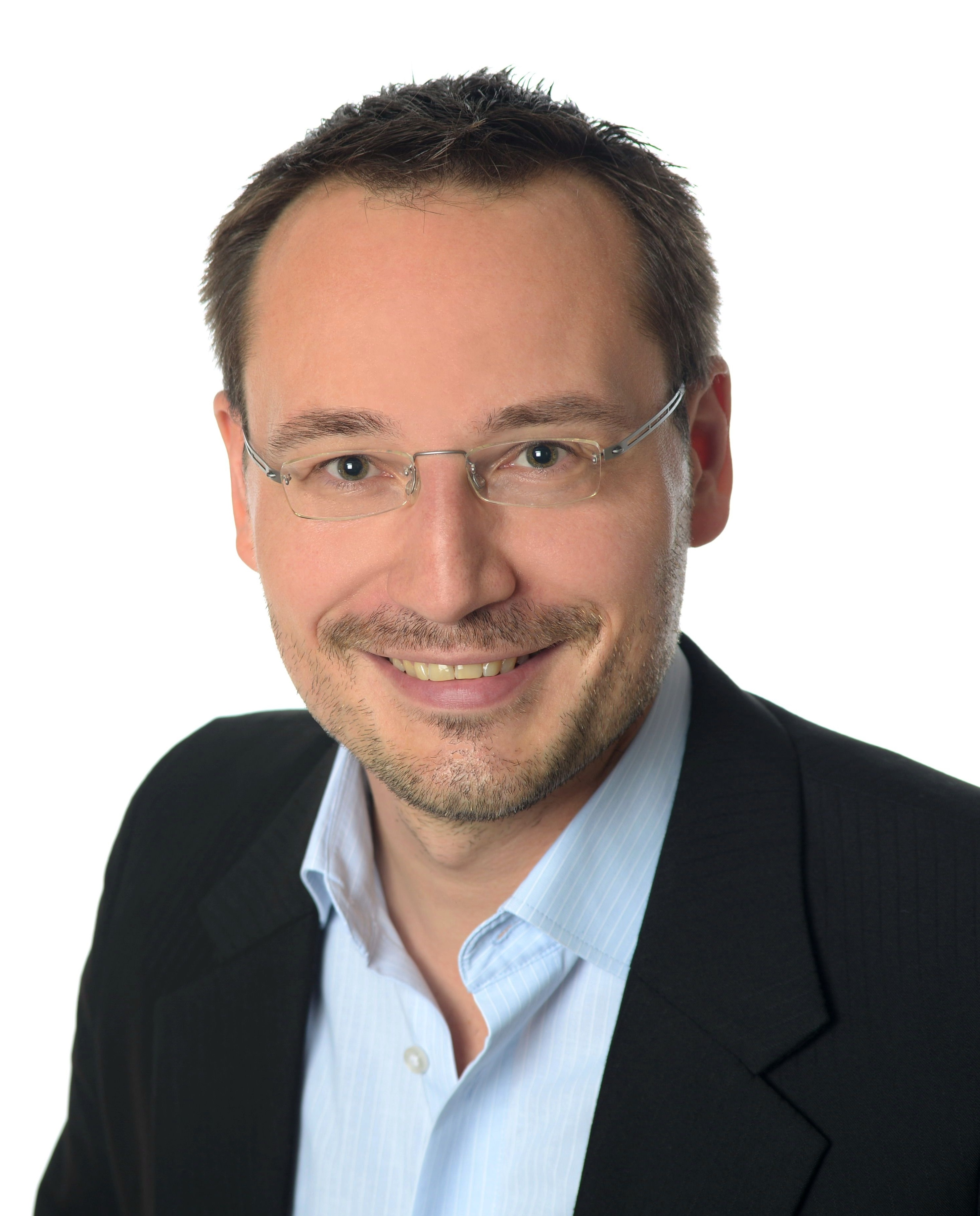}}]{Christian Janiesch} is assistant professor with the Julius-Maximilians-Universit\"at W\"urzburg.
His research is at the intersection of business process management and business analytics with frequent applications in the Industrial Internet of Things. He is on the Department Editorial Board for BISE and has authored over 100 scholarly publications. His work has appeared in prestigious journals 
as well as in the major international conferences in the area.
\end{IEEEbiography}

\begin{IEEEbiography}[{\includegraphics[width=1in,height=1.25in,clip,keepaspectratio]{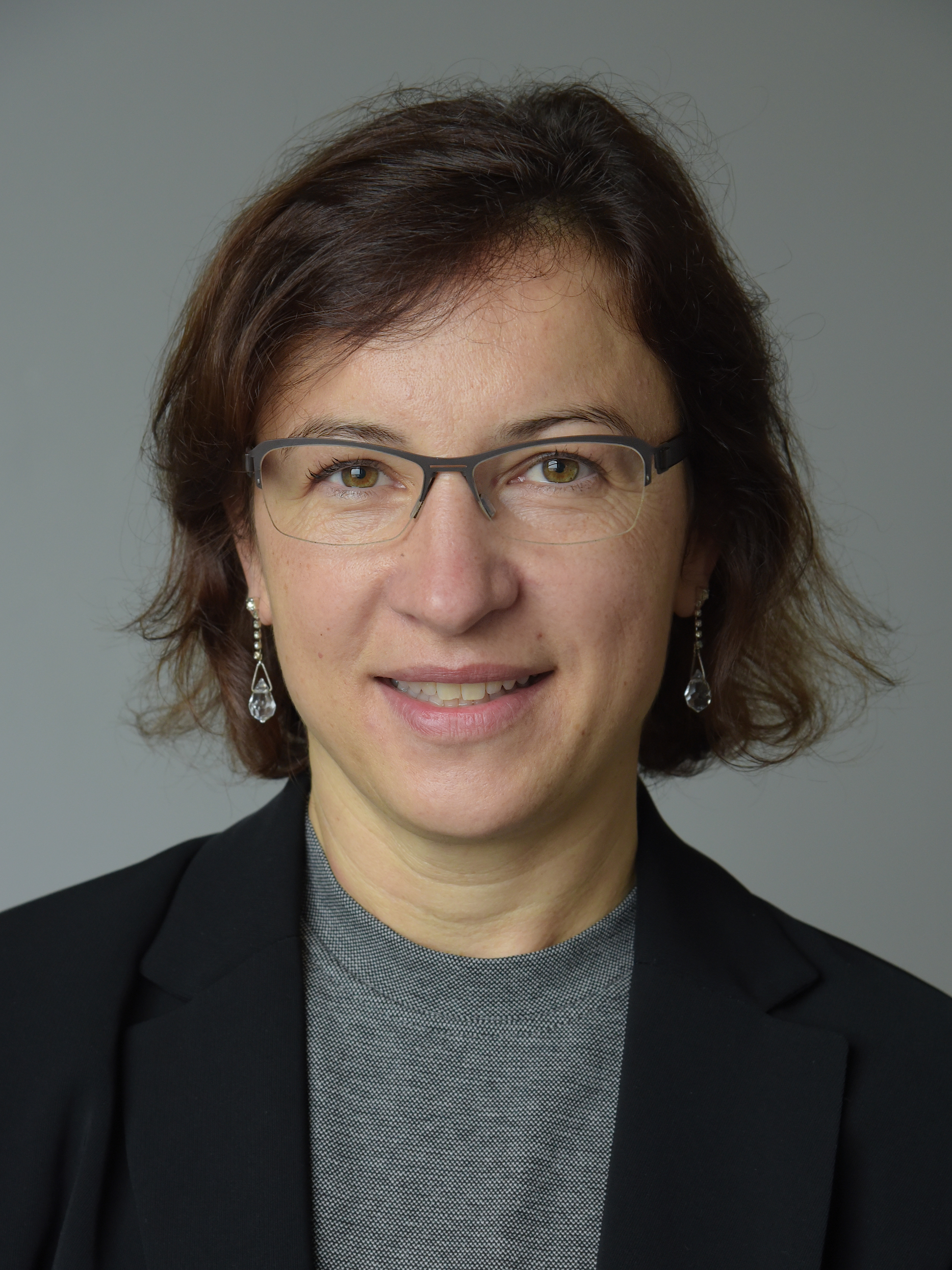}}]{Agnes Koschmider} is a professor of information systems at the Kiel University and head of the Process Analytics group. Her research interest broadly explores how to extract meaningful event logs from IoT (information) systems with special focus on privacy preservation. Her work has been published in over 90 research papers and articles. \end{IEEEbiography}

\begin{IEEEbiography}[{\includegraphics[width=1in,height=1.25in,clip,keepaspectratio]{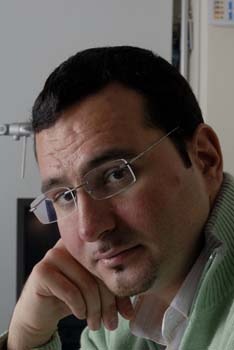}}]{Massimo Mecella} is an associate professor 
	with Sapienza Universit\`a di Roma. His research focuses on business process management, cyber-physical systems and Internet-of-Things, advanced interfaces and human-computer interaction, software architectures and service-oriented computing, with applications in multiple fields including digital government, smart spaces, Industry 4.0, healthcare, disaster/crisis response \& management, cyber-security, digital humanities. 
	He published over 180 research papers and chaired different conferences in the above areas.
\end{IEEEbiography}

\begin{IEEEbiography}[{\includegraphics[width=1in,height=1.25in,clip,keepaspectratio]{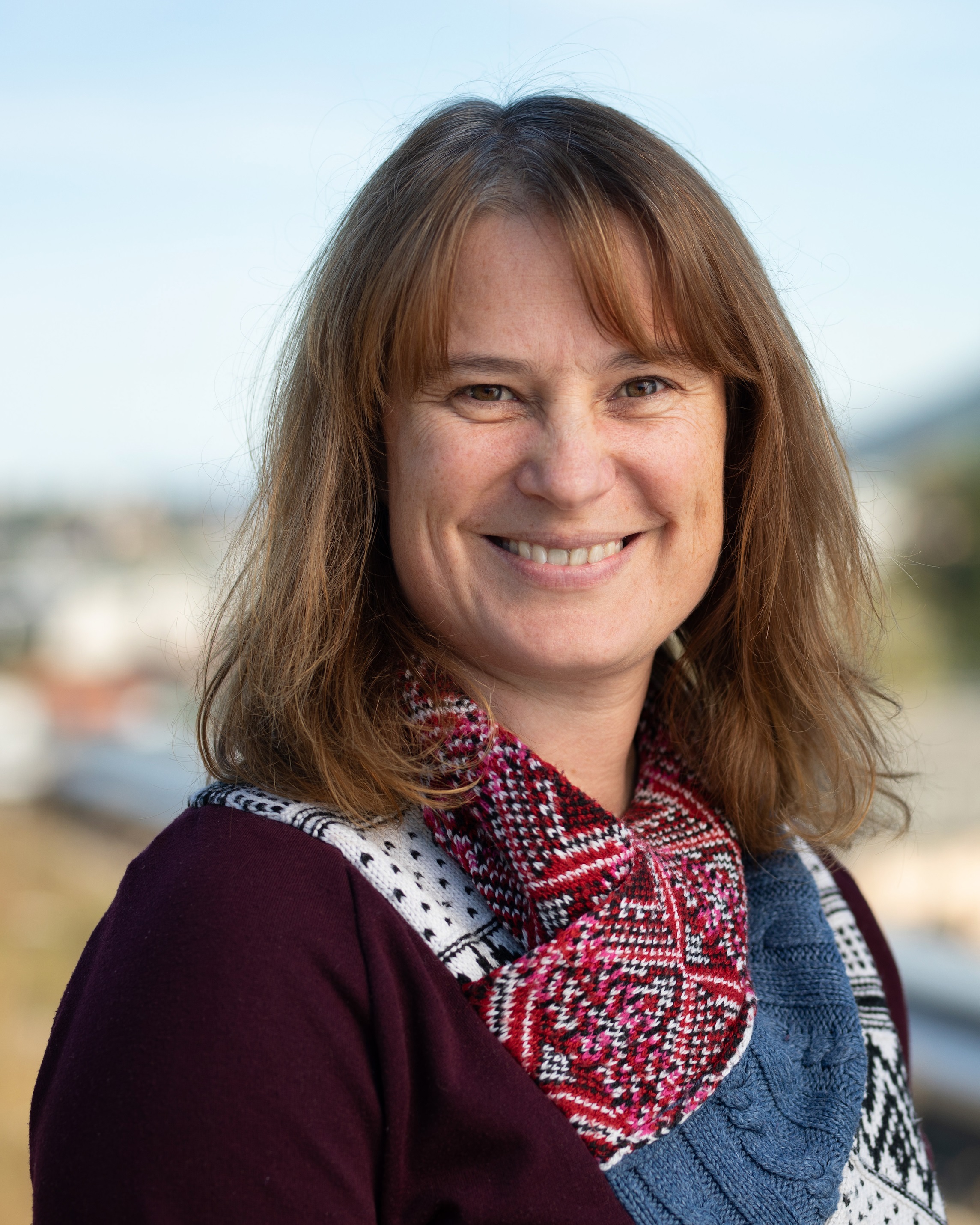}}]{Barbara Weber} is Chair for Software Systems Programming and Development and Director of the Institute of Computer Science at the University of St. Gallen, Switzerland. Barbara holds a PhD in business administration from the University of Innsbruck, Austria.
Her research interests include the development of adaptive software systems including flexible and adaptive business process support and neuro-adaptive information systems as well as human and cognitive aspects of software and information systems engineering. She authored more than 160 research papers and articles.
\end{IEEEbiography}

\begin{IEEEbiography}[{\includegraphics[width=1in,height=1.25in,clip,keepaspectratio]{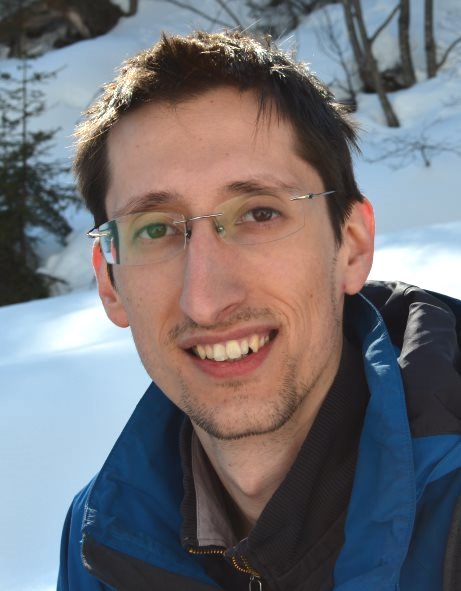}}]{Andrea Burattin} is associate professor at the Technical University of Denmark.
He obtained his Ph.D. degree from the Universities of Bologna and Padua (Italy). The IEEE Task Force on Process Mining awarded to his Ph.D. thesis the Best Process Mining Dissertation Award for theses defended in 2012-2013. A reworked version of his thesis has then been published as Springer Monograph in the LNBIP series.
Since 2019 he is a member of the the Steering Committee of the IEEE Task Force on Process Mining.\end{IEEEbiography}

\begin{IEEEbiography}[{\includegraphics[width=1in,height=1.25in,clip,keepaspectratio]{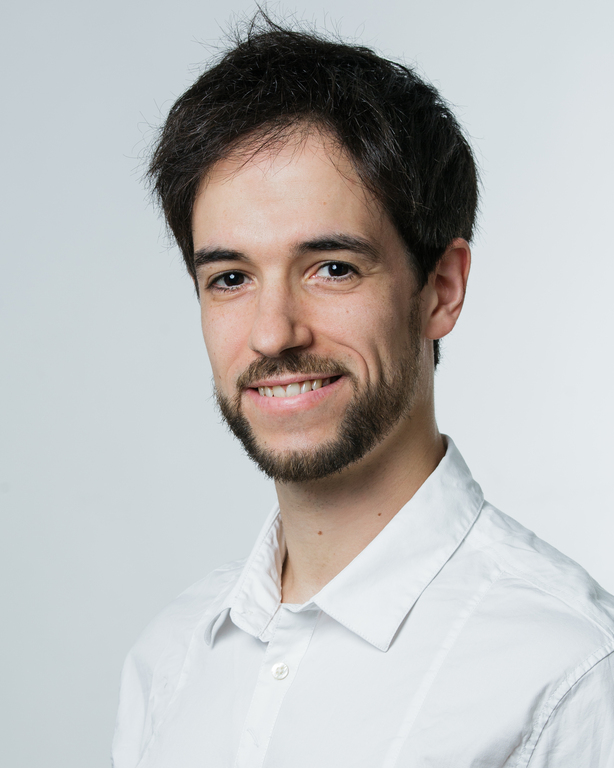}}]{Claudio Di Ciccio} is an assistant professor with Sapienza Universit\`a di Roma, Italy. His research interests include process mining, declarative modelling, and blockchains. His has published more than 70 papers, in the proceedings of renowned international conferences such as BPM and CAiSE, and top journals.
He was PC chair of the first Blockchain Forum at BPM 2019 and regularly serves on the Program Committee of the conferences in the area. He is a member of the Steering Committee of the IEEE Task Force on Process Mining. 
\end{IEEEbiography}

\begin{IEEEbiography}[{\includegraphics[width=1in,height=1.25in,clip,keepaspectratio]{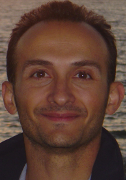}}]{Giancarlo Fortino} is full professor at University of Calabria (Unical), Italy.
He is the director of the SPEME lab at Unical as well as co-chair of Joint labs on IoT established between Unical and WUT and SMU and HZAU chinese universities, respectively. His research interests include agent-based computing, wireless (body) sensor networks, and IoT. He is author of 450+ papers in international l journals, conferences and books. He is cofounder and CEO of SenSysCal S.r.l., a Unical spinoff focused on innovative IoT systems.
\end{IEEEbiography}

\begin{IEEEbiography}[{\includegraphics[width=1in,height=1.25in,clip,keepaspectratio]{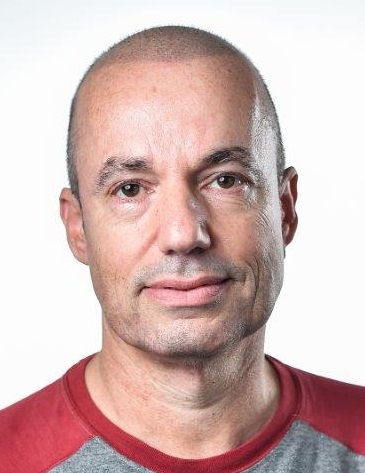}}]{Avigdor Gal} is a full professor of Data Science
	with Technion -- Israel Institute of Technology. His research focuses on data integration, business process management, cyber-physical systems and Internet-of-Things, with applications in multiple fields including smart cities and food IoT.
	He published more than 170 research papers in leading journals
	and conferences
	and served as a program and general chair of BPM and DEBS. Avigdor is the author of the ``Uncertain Schema Matching'' book.
\end{IEEEbiography}

\begin{IEEEbiography}[{\includegraphics[width=1in,height=1.25in,clip,keepaspectratio]{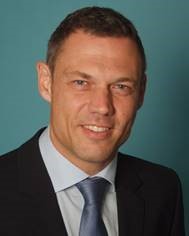}}] {Udo Kannengiesser} is a professor at Johannes Kepler University Linz, Austria. His research focuses on design computing, design cognition, design creativity, design methodology, distributed systems, process management, and digital manufacturing. He previously worked for several software companies, research institutes and universities in Australia, Germany and Austria.
\end{IEEEbiography}

\begin{IEEEbiography}[{\includegraphics[width=1in,height=1.25in,clip,keepaspectratio]{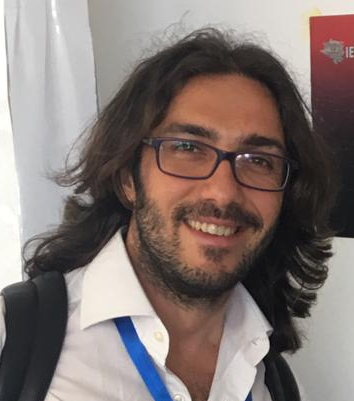}}] {Francesco Leotta} is assistant professor at Department of Computer, Control and Management Engineering of Sapienza Universit\`a di Roma, Italy. His research concerns algorithmic, methodological, experimental and practical aspects in different areas of Computer Science, including ubiquitous computing, human-computer interaction and digital humanities. Such topics are challenged in the application domains of smart spaces, smart manufacturing and cultural heritage.
\end{IEEEbiography}

\begin{IEEEbiography}[{\includegraphics[width=1in,height=1.25in,clip,keepaspectratio]{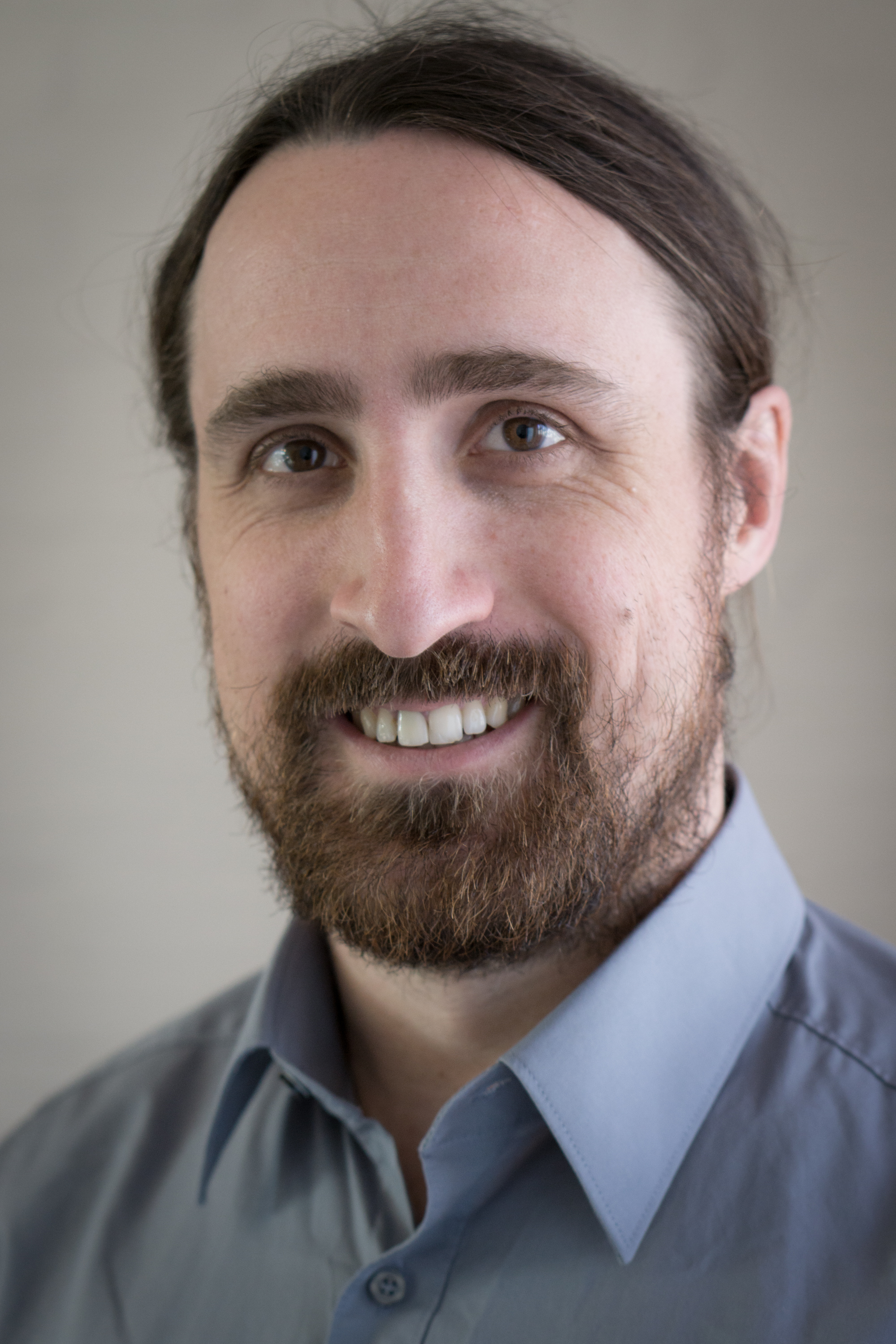}}]{Felix Mannhardt} is a part-time associate professor with the NTNU Norwegian University of Science and Technology in Trondheim, Norway. He also works as research scientist at the Technology Management department of SINTEF Digital where he co-founded KIT-AR Ltd.
His research interests include how to use low-level event data for process mining
as well trust and privacy concerns connected to process mining. 
He is a member of the IEEE Task Force on Process Mining and contributed significantly to the frameworks ProM and bupaR. His doctoral thesis was awarded the Process Mining Dissertation Award at the ICPM 2019.
	
\end{IEEEbiography}

\begin{IEEEbiography}[{\includegraphics[width=1in,height=1.25in,clip,keepaspectratio]{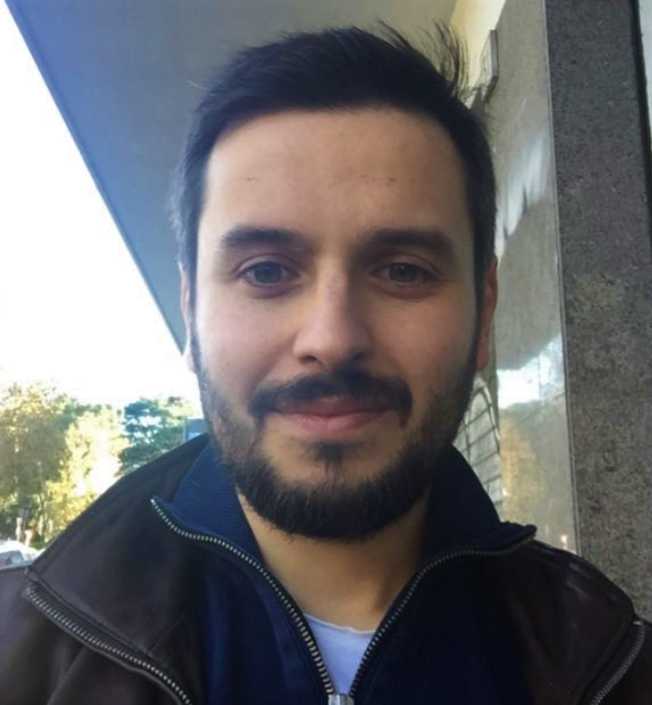}}]{Andrea Marrella}  is an assistant professor at Sapienza Universit\`a di Roma, Italy. His research activity focuses on how to integrate Artificial Intelligence with Business Process Management solutions, to untangle complex challenges from the fields of Process Mining and Robotic Process Automation. He has co-authored more than 70 peer-reviewed publications in renowned international conferences
and top journals.
Since 2017, he is the Information Director of the ACM Journal on Data Quality.\end{IEEEbiography}

\begin{IEEEbiography}[{\includegraphics[width=1in,height=1.25in,clip,keepaspectratio]{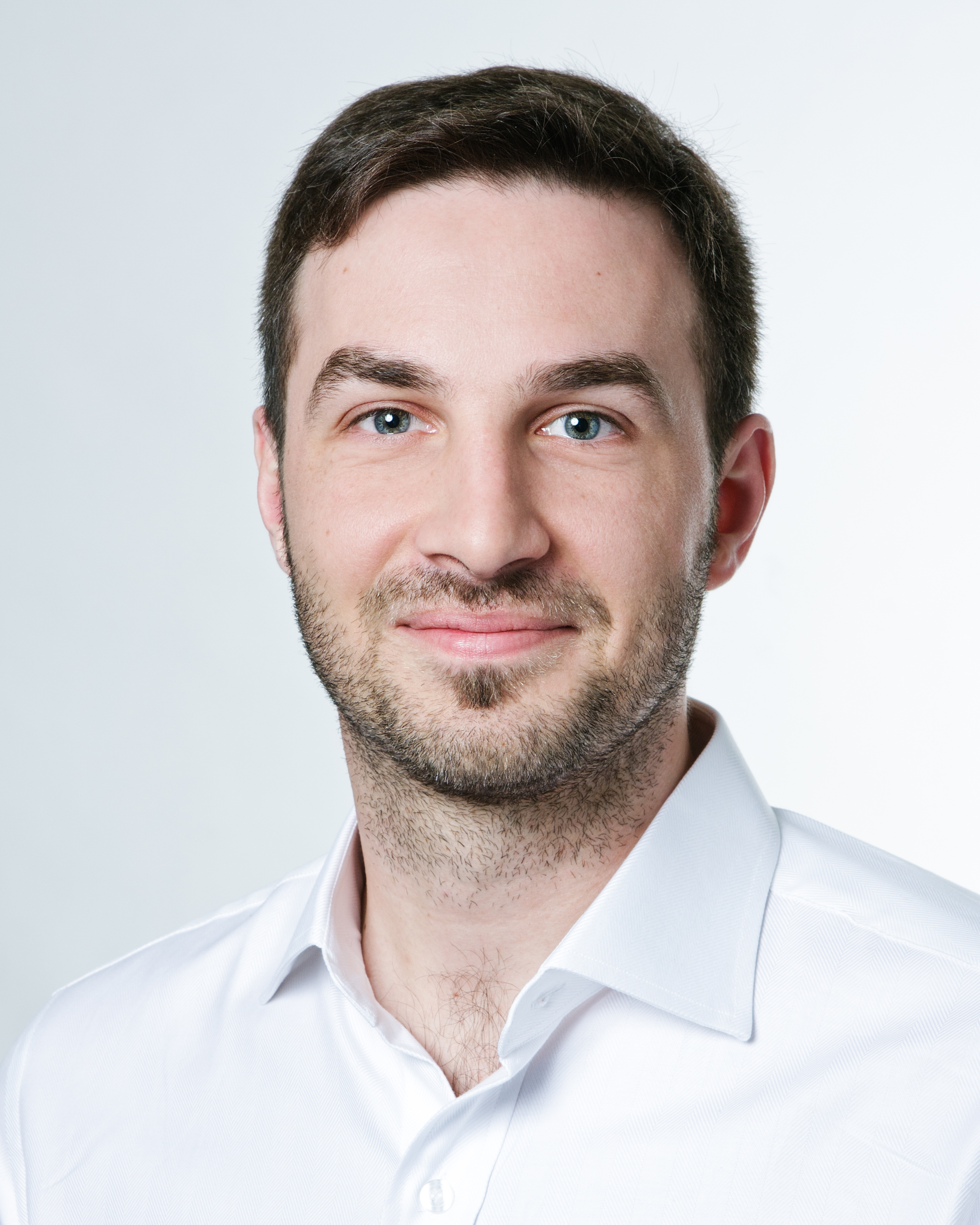}}]{Jan Mendling} is a full professor at Wirtschaftsuniversit{\"a}t Wien (WU Vienna), Austria. His research interests include business process management and information systems. He is co-author of the textbooks Fundamentals of Business Process Management (http://fundamentals-of-bpm.org/) and Wirtschaftsinformatik. He has published more than 400 research papers and articles. He is member of several international journals, member of the board of the Austrian Society for Process Management (http://prozesse.at), a co-founder of the Berlin BPM Community of Practice (http://www.bpmb.de), and member of the IEEE Task Force on Process Mining.
\end{IEEEbiography}

\begin{IEEEbiography}[{\includegraphics[width=1in,height=1.25in,clip,keepaspectratio]{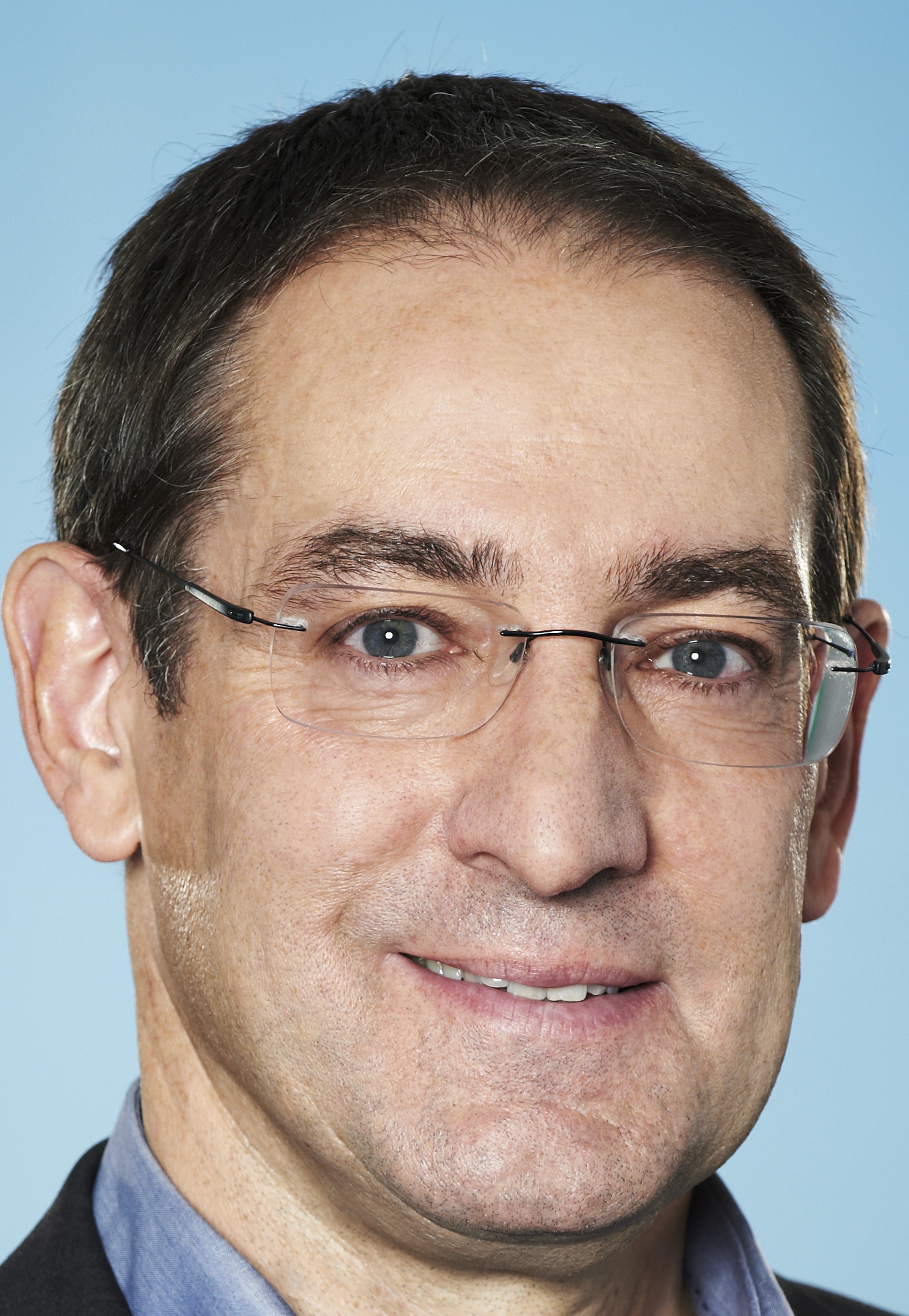}}]{Andreas Oberweis} is professor for Applied Informatics at the Karlsruhe Institute of Technology (KIT) since 2003. From 1995 to 2003 he was professor for Information Systems Development at Goethe-University Frankfurt/Main. His research and teaching activities are at the borderline between software engineering, BPM, and database management. He is member of the board of FZI Research Center for Information Technology in Karlsruhe and he is co-founder of several companies in the field of business process management.
	
\end{IEEEbiography}

\begin{IEEEbiography}[{\includegraphics[width=1in,height=1.25in,clip,keepaspectratio]{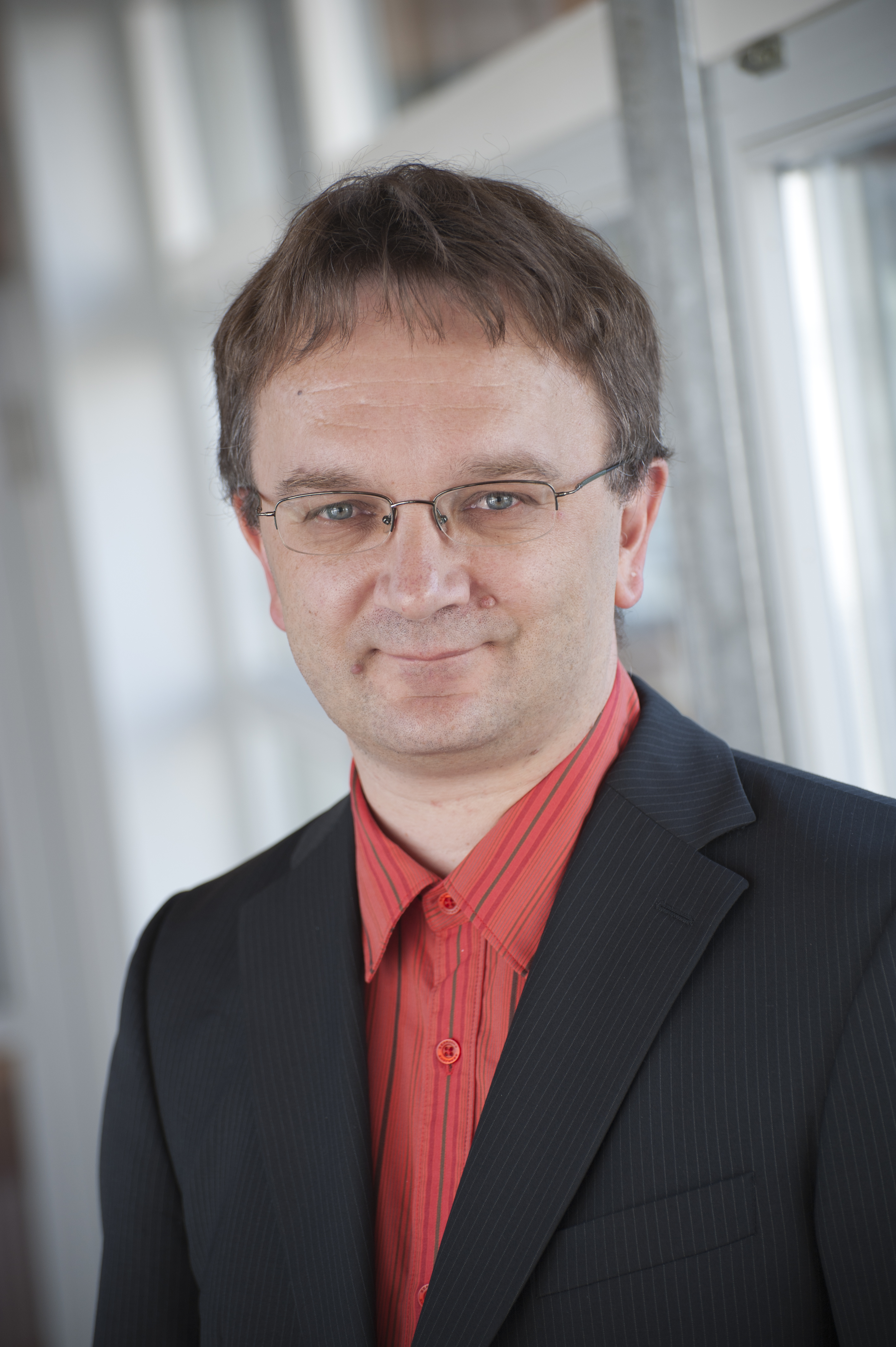}}]{Manfred Reichert} is professor of Computer Science and Director of the Databases and Information Systems Institute at Ulm University, Germany. His research spans across the fields of process and data science, information systems engineering, and digital services. Several of his techniques have been applied in healthcare, logistics, automotive engineering, and Industry 4.0.
He published more than 500 refereed articles and received several best paper awards (e.g. CoopIS'05, EDOC'08, EDOC'18, AIMS'17, BPM Test of Time Award 2013). Finally, he is co-founder of the AristaFlow Ltd. and co-author of a Springer book on process flexibility.
\end{IEEEbiography}

\begin{IEEEbiography}[{\includegraphics[width=1in,height=1.25in,clip,keepaspectratio]{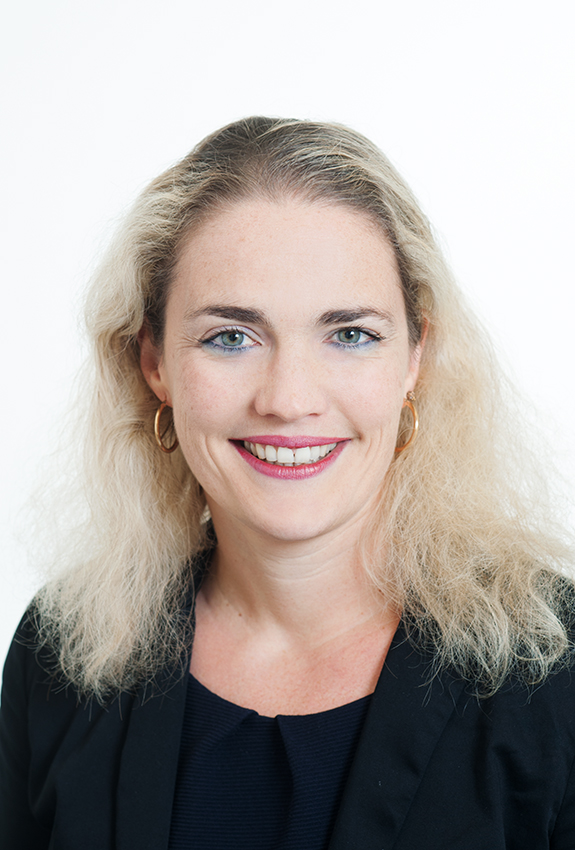}}]{Stefanie Rinderle-Ma} is full professor at the Faculty of Computer Science, University of Vienna, Austria. Her research focuses on process-aware information systems, distributed process technology, digitalized compliance management, and business process intelligence with application areas such as manufacturing and health care. Both pose high demands on business processes and IoT. Manufacturing processes equipped with sensor data, for example, have been implemented and analyzed in cooperation with the Austrian Center for Digital Production. She has published more than 200 articles and conference papers.
\end{IEEEbiography}

\begin{IEEEbiography}[{\includegraphics[width=1in,height=1.25in,clip,keepaspectratio]{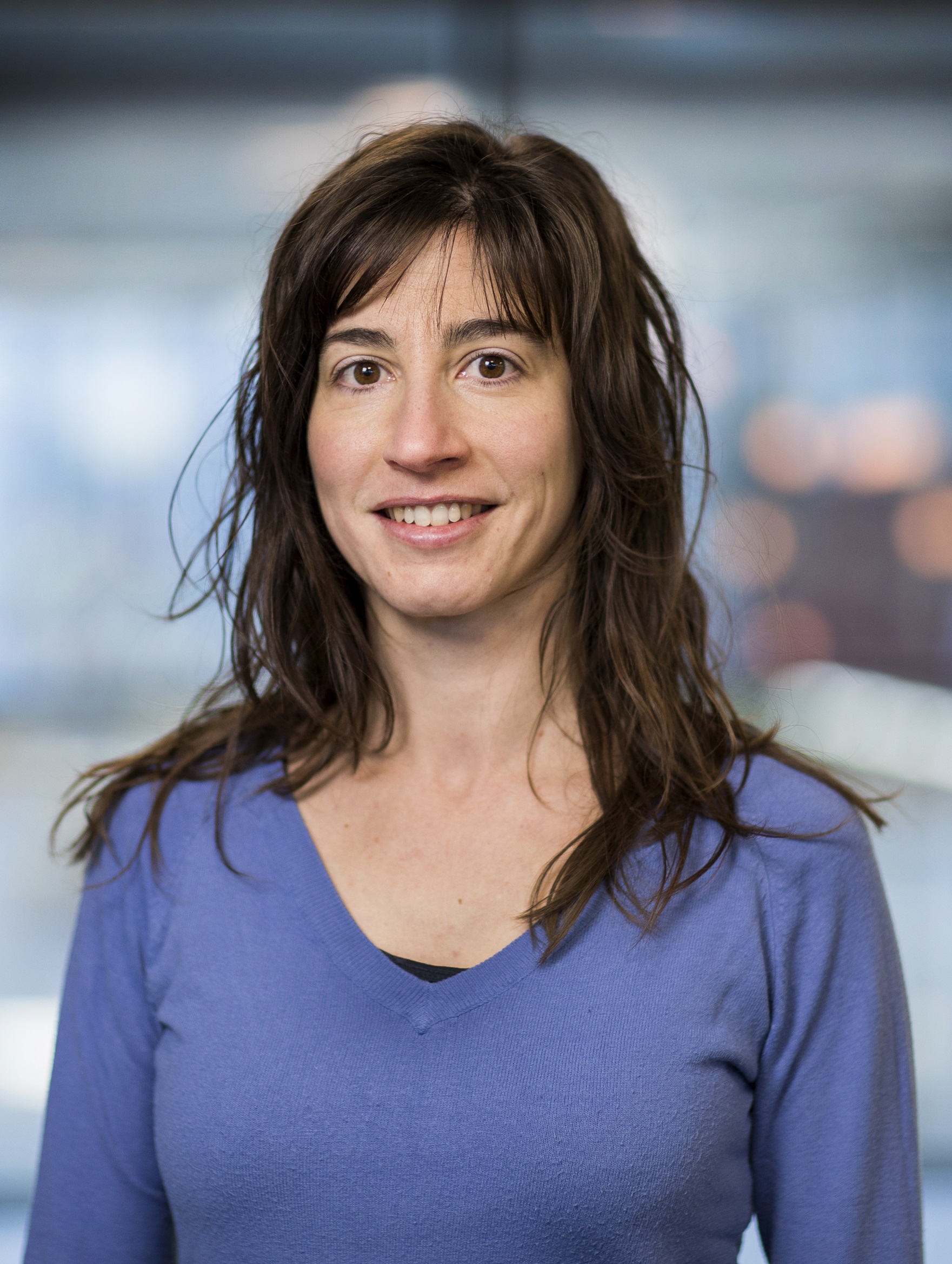}}]{Estefan\'{i}a Serral} is an assistant professor at KU Leuven, Belgium. Her research focuses on Internet of Things, Business Process Management, and context-adaptive systems. In 2018, she was also an Assistant professor at TU/e, The Netherlands. From 2012 to 2014, she led the Semantic Knowledge Representation and Integration research group at the CDL-Lab at the Technical University of Vienna, Austria. Until 2012, she worked in the ProS Research Center at the Technical University of Valencia (Spain), where she designed a novel method for developing IoT systems using Model-Driven Development (MDD) and Semantic technologies. She has authored more than 100 research papers in conferences and journals.
\end{IEEEbiography}

\begin{IEEEbiography}[{\includegraphics[width=1in,height=1.25in,clip,keepaspectratio]{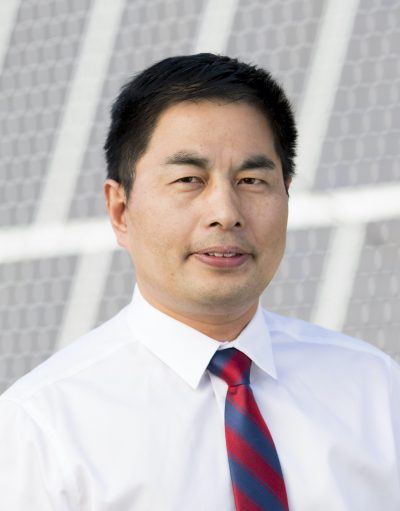}}]{WenZhan Song} is a Chair professor in Electrical and Computer Engineering with University of Georgia. His research focuses on cyber-physical systems informatics and security and their applications in energy, health and environment systems, where distributed sensing, networking, computing and security play a critical role and need a transformative study..
	He published near 200 research papers and chaired different conferences in the above areas and received numerous awards including NSF CAREER Award 2010 and Mark Weiser Best Paper Award 2020.
\end{IEEEbiography}

\begin{IEEEbiography}[{\includegraphics[width=1in,height=1.25in,clip,keepaspectratio]{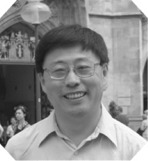}}]{Jianwen Su} is professor  of Computer Science
	at the University of California, Santa Barbara. His research focuses on data management and business process management, Internet-of-Things, and service-oriented computing.  He published many research papers and chaired different conferences in the above areas, including PODS, SIGMOD, and ICSOC. 
\end{IEEEbiography}

\begin{IEEEbiography}[
	
	{\includegraphics[width=1in,height=1.25in,clip,keepaspectratio]{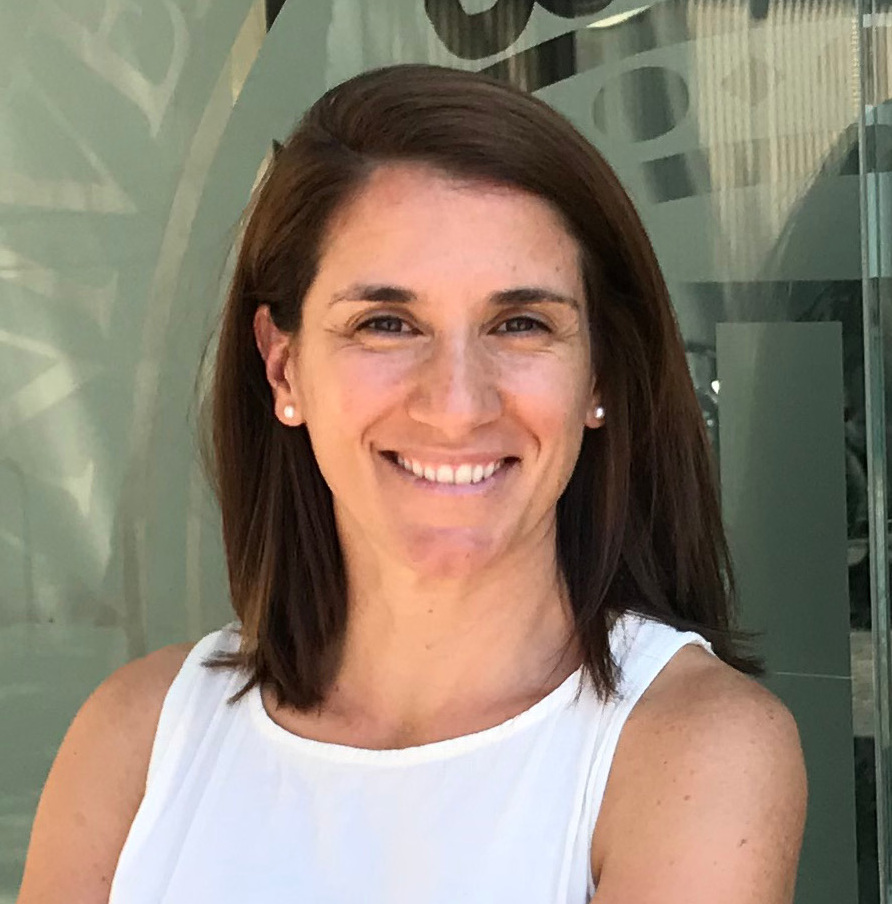}}]{Victoria Torres} is associate professor with Universitat Polit\`ecnica de Val\`encia. She is a member of the Valencian Research Institute for Artificial Intelligence (VRAIN), working actively in areas such as the Business Process Management, Internet of Things, Metamodeling, Web Engineering, and Model Driven Development. She has published several papers in conferences and journals such as Information and Software Technology, Information Systems, and Software and System Modeling. Since 2010, she is actively participating in National and European Projects (OPEES, SITAC, DECODER).\end{IEEEbiography}

\begin{IEEEbiography}[{\includegraphics[width=1in,height=1.25in,clip,keepaspectratio]{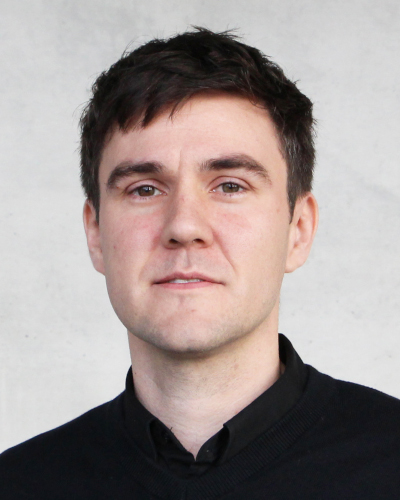}}]
{Matthias Weidlich} is professor and Chair of Databases and Information
Systems at the Department of Computer Science at Humboldt-Universit\"at
zu Berlin. His research focuses on process-oriented and event-based
information systems, and his results appear regularly in premier
conferences (SIGMOD, VLDB, ICDE, IJCAI, BPM) and journals (TKDE, Inf.
Sys., VLDB Journal) in the field. He is a Junior-Fellow of the German
Informatics Society and in 2016 received the Berlin Young Researcher
Award. He serves as editor-in-chief for the Information Systems journal
and is a member of the steering committee of the ACM DEBS conference series.
\end{IEEEbiography}

\begin{IEEEbiography}[{\includegraphics[width=1in,height=1.25in,clip,keepaspectratio]{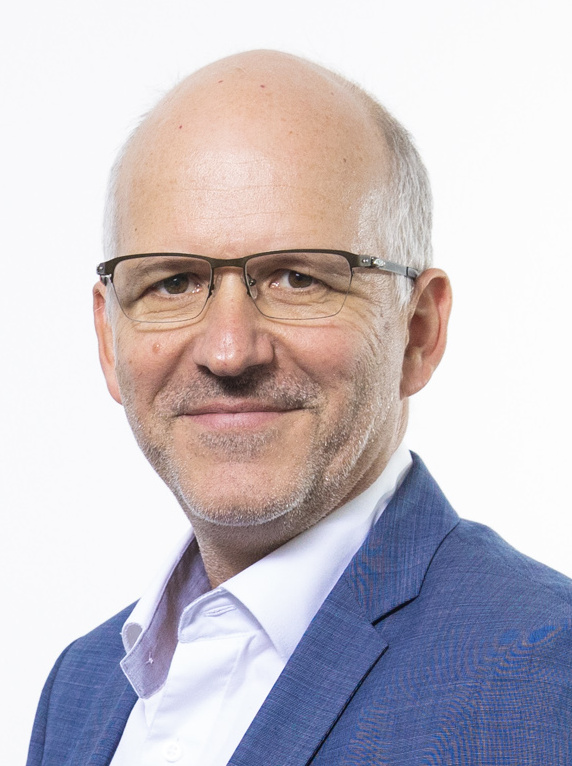}}]{Mathias Weske} is professor and chair of the business process technology research group at Hasso Plattner Institute at the Digital Engineering Faculty, University of Potsdam, Germany. His research focuses on the engineering of process oriented information systems, process mining, and event processing. Dr. Weske is author of the first textbook on business process management and he held the first massive open online course on the topic in 2013. He is on the Editorial Board of Springer’s Distributed and Parallel Databases journal, Springer’s Computing journal, and he is a founding member of the steering committee of the BPM conference series and, since September 2017, chair of the steering committee.
\end{IEEEbiography}
\enlargethispage{-5in}

\begin{IEEEbiography}[{\includegraphics[width=1in,height=1.25in,clip,keepaspectratio]{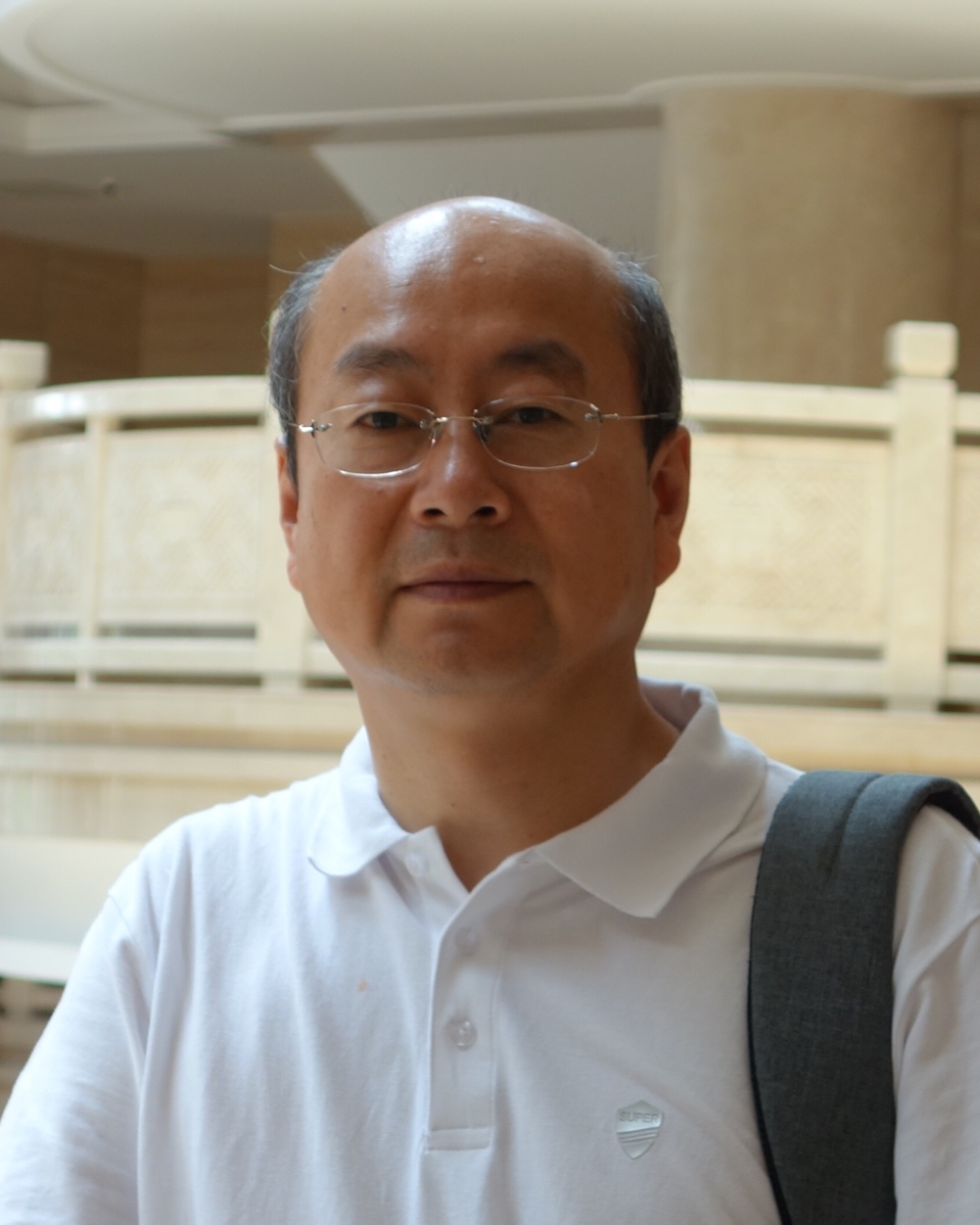}}]{Liang Zhang} is a full professor of computer science at Fudan University, China.  His current research interests include IoT-enabled information systems, the reactive systems, mainly in the form of service-based ones or business process-oriented. He authored more than 100 research papers. \end{IEEEbiography}






\end{document}